\documentclass[12pt]{article}

\usepackage[T1,OT1,TS1,T2A]{fontenc}
\usepackage{epsfig,amsmath,amsfonts,amssymb}

\usepackage{ifthen}

\def \beq {\begin{equation}}
\def \eeq {\end{equation}}
\def \beqar {\begin{eqnarray}}
\def \eeqar {\end{eqnarray}}
\def \beqa* {\begin{eqnarray*}}
\def \eeqa* {\end{eqnarray*}}
\def \pa {\partial}

\def \d {\,{\rm{d}}}

\def\refart#1#2#3#4#5#6#7{{#1}, \textit{#3} \textbf{#4} {(#6)} {#5}{\ifthenelse{\equal{#7}{}}{}{ [#7]}}}%

\def\refbook#1#2#3#4#5#6#7{{#1}, \textit{#2}, #3, {\ifthenelse{\equal{#4}{}}{}{ch.~#4}}, #5, #6, #7}%

\begin{document}
\title{Nonsingular Chaplygin gas cosmologies \\in universes connected by wormhole}%
\author{Anna~Mokeeva$^{1)}{}^\dag$ and Vladimir~Popov$^{2)}{}^\ddag$}%
\date{}

\maketitle%

{\small
\noindent$^{1)}$\textit{Physical Department, Lobachevsky State University of Nizhni Novgorod, Gagarin ave. 23, Nizhni Novgorod 603950, Russia}\\
$^{2)}$\textit{Institute of Physics, Kazan Federal University, Kremlyovskaya st. 18, Kazan 420008, Russia}\\
\\
$^\dag$anne.zaharovoy@gmail.com\\
$^\ddag$vladipopov@mail.ru
}

\begin{abstract}
We present some exact solutions of the Einstein equations with an
anisotropic fluid exploiting the Chaplygin equation of state. The
solutions describe spacetimes with two identical T~regions and an
intermediate static spherically symmetric R~region containing a
wormhole. The metric in the T~region represents an anisotropic
Kantowski-Sachs cosmological model.  Its evolution starts from a
horizon and develops according to different scenarios including
eternal expansion, contraction and also a finite universe lifetime.
\end{abstract}

\vspace*{12pt}

{\small

}

\newpage

\section{Introduction}

Observational data of the last decade,  such as Type Ia Supernovae
(SNIa)~\cite{Riess&Perlmutter}, Cosmic Microwave Background
(CMB)~\cite{BOOMERanG&WMAP} and Large Scale Structure~\cite{SDSS},
are evidence of accelerating expansion of the Universe. This
behavior is caused by the dark energy featured by a negative ratio
of the pressure to the energy density. A dark energy nature is not
clear nowadays and a number of competing models describe the dark
energy as an exotic matter or as a result of non-minimal
gravitational interaction with ordinary matter.

The simplest dark energy model, the cosmological constant, is indeed
the vacuum energy with the equation of state $p=-\rho$. A number of
models, such as quintessence \cite{Wetterich1}, k-essence
\cite{Armendariz1}, phantom \cite{Caldwell1} and etc., are based on
scalar field theories. An alternative way is to be thought a
geometric origin of the dark energy. It motivates to modify the
four-dimensional gravity theory \cite{Nojiri1} or to introduce
additional dimensions as in braneworld models~\cite{braneworld}. The
Chaplygin gas model, also denoted as quartessence, exploits a
negative pressure fluid, which is inversely proportional to the
energy density $p=-\alpha/\rho$ \cite{Kamenshchik1}. It was modified
for the models unifying the dark energy and dark matter
\cite{Debnath1}, including
generalized Chaplygin gas \cite{Bento&Bilic,Bento2} and superfluid
Chaplygin gas \cite{SCG 1}.  For a more detail review of dark
energy models and references see \cite{Copeland1}.

In the cosmological context the Chaplygin  gas originally appeared
as a hydrodynamical representation of the generalized Born-Infeld
Lagrangian
\beq
{\cal L}_\text{BI} = -\sqrt\alpha \sqrt{1-\pa_\nu \theta \, \pa^\nu \theta}
\eeq
describing a (3+1)-dimensional brane universe with the scalar
field $\theta$ in a (4+1)-dimensional bulk \cite{Bento&Bilic}. It
was also shown that the Chaplygin gas model can be derived from
the Lagrangian \beq\label{LagrangianBasic} {\cal L} = \pa_\nu
\phi^* \pa^\nu \phi -M \left( \frac{\phi^*\phi}{\lambda} +
\frac{\lambda}{\phi^*\phi} \right)
\eeq
for a complex scalar field $\phi$ in the WKB-approximation \cite{SCG 1}.

The Chaplygin equation of state can violate the strong energy
condition  $\rho+3p>0$, which is necessary for an accelerating
cosmological process as well as it can violate  the null energy
condition $\rho+p>0$ providing for the existence of traversable
wormholes. Wormholes with the Chaplygin gas are originally studied
by Lobo~\cite{Lobo1}. Owing to the peculiar equation of state the
obtained solutions possess some features. Primarily, if the solution
exists in the whole spacetime then it cannot be asymptotically flat.
To ensure  asymptotic flatness for the Chaplygin wormhole the
interior solution in \cite{Lobo1}  is matched with the exterior
Schwarzschild solution using the Darmois-Israel formalism of a
dynamic thin shell \cite{Darmois}. This method is also used when the
wormhole solution exists only within a spatial sphere and cannot be
directly continued outside because it is singular on the sphere.

An alternative approach to the Chaplygin wormholes is to consider
horizons instead of singularities. A cosmic expansion beyond an
event horizon can be considered as a model of an origin of our
Universe \cite{Pathria,Frolov,Bronnikov1,Bronnikov2}. The idea that
the universe can emerge in the interior of a black hole was proposed
by Pathria \cite{Pathria}. The primary attention to the universes
inside the black holes is concentrated on the inquiry of geometrical
properties and cosmological details rather than on exact solutions
of the Einstein equations \cite{Frolov}.  Bronnikov et al.
\cite{Bronnikov1} used the exact solutions for phantom scalar fields
to study and classify regular black holes where singularity is
replaced by a cosmological expansion and which therefore were called
\emph{black universes}. Such objects can be supported by phantom
matters and could arise from collapse in another ambient universe.
In the work \cite{Bronnikov2} the Lema\^{\i}tre type cosmology for
an anisotropic perfect fluid with the vacuum equation of state was
considered for different topologies of the spacetime. The conditions
for the cosmological evolution starting from a horizon for different
kinds of matter were studied in \cite{Bronnikov3,Bronnikov4}.

We apply the similar structure to Chaplygin wormholes. It is free of
singularities and the wormhole is surrounded by horizons separating
the spacetime into the static and cosmological regions. In this work
we deal with the simple horizons for which we determine the
conditions of regularity for the metric and the Chaplygin matter.
Some exact solutions corresponding to different cosmological
scenarios are developed using these conditions.

The paper is organized as follows.  In the next section we consider
the general properties of spacetimes with a spherically symmetric
wormhole surrounded by two horizons and external universes which are
supported by matter with the Chaplygin equation of state. Some
particular cosmological scenarios are presented in
Sec.~\ref{Sec-ExactSolutions}. Concluding remarks are summarized in
Sec.~\ref{Sec-Conclusion}.

\section{Wormhole between two horizons}

We consider a static spherically symmetric spacetime with the
metric
\begin{equation} \label{SSM}
\d s^2=A(r)\d t^2-\frac{\d r^2}{1-b(r)/r}-r^2(\d\theta^2+\sin^2\theta \d\varphi^2)
\end{equation}
and the stress-energy tensor of an anisotropic perfect fluid in
which the energy density and the radial pressure are related by
Chaplygin's equation of state
\begin{equation} \label{Chap}
 p_{r}=-\dfrac{\alpha}{\rho},
\end{equation}
where $\alpha$ is a positive constant.
In this case the Einstein equations $G_{\mu\nu}=8\pi T_{\mu\nu}$ (we set 
$c=G=1$) reduce to
\beqar
\label{Ein1} b' &=& 8\pi r^2\rho,
\\
\label{Ein2} \frac{A'}{A} &=& \dfrac{b+8\pi r^3p_r}{r^2(1-b/r)},
\\
\label{Ein3} p'_r &=& \dfrac{2}{r}(p_{t}-p_r)-(\rho+p_r)\frac{A'}{A},
\eeqar
where the prime denotes the derivative with respect to $r$ and $p_{t}(r)$ is the transversal pressure.

The metric (\ref{SSM}) describes a static spherically symmetric
wormhole if the radial coordinate $r$ increases from a minimum
value, corresponding to the wormhole throat, to infinity. Without
loss in generality we take $r_\text{min}=1$. In this approach $r$ is
associated with two maps covering the spacetime on either side of
the throat. The functions $A(r)$ and $b(r)$ are regular and defined
through all the range $1\le r<\infty$ and satisfy the flare-out
conditions
\beqar
& \label{throatCond1} b(1)=1,\qquad b'(1)<1 &\\&
\label{throatCond2} 0<A(1) <\infty,\qquad A'(1)<\infty &
\eeqar
to provide a minimal area  for coordinate spheres at $r=1$ and the
metric regularity on the throat.

For traversable wormholes it is also necessary that $A(r)$ should be
positive and $b(r)<r$ elsewhere to avoid horizons and
singularities. These requirements are ruled out in the present paper in favor of the horizon located at the radius $r=r_0>1$.

\begin{figure}[t]
\includegraphics[width=.45\textwidth]{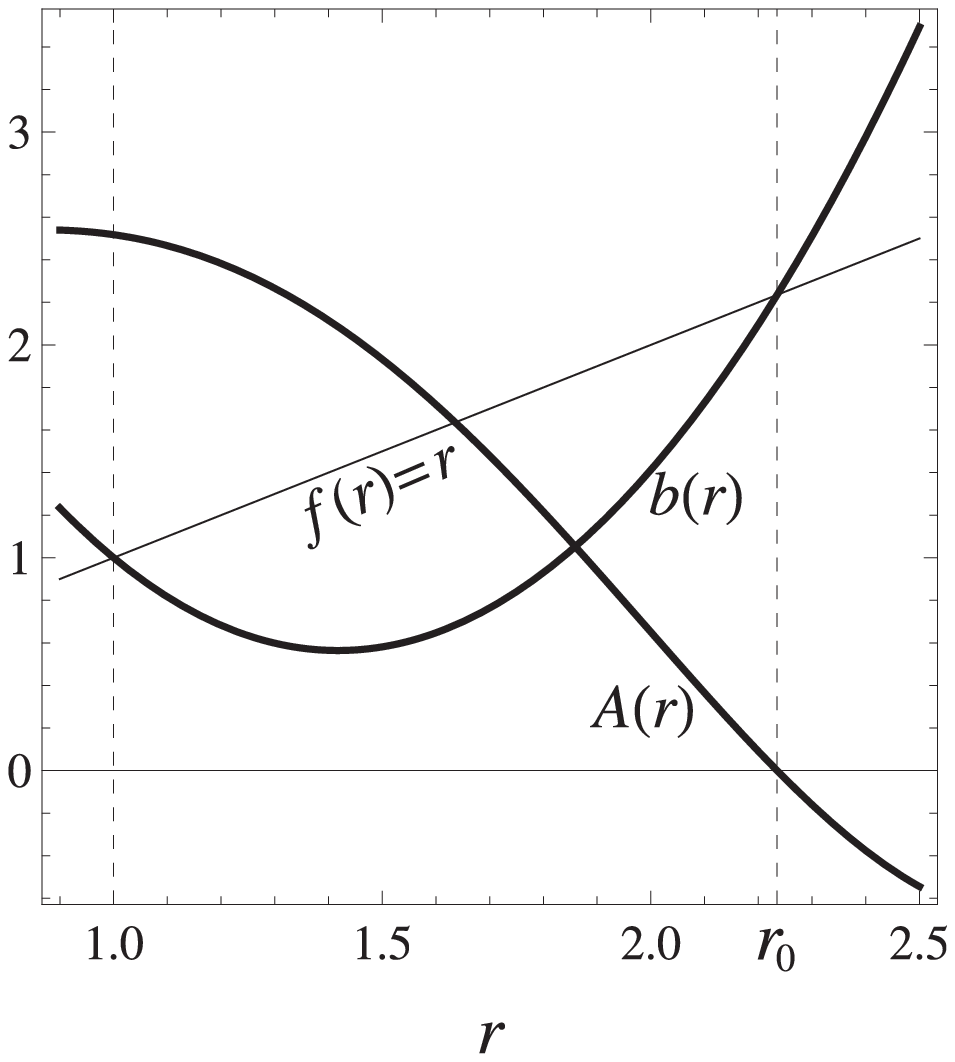}\hfill%
\raisebox{.8\height}{\includegraphics[width=.51\textwidth]{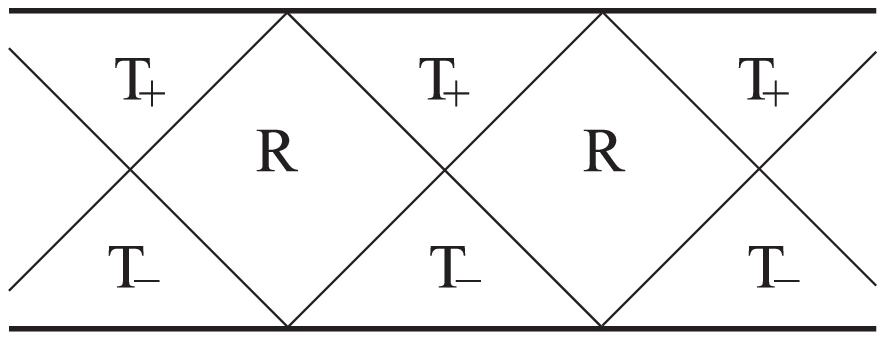}}
\\%
\parbox[t]{0.45\textwidth}{\caption{The functions $A(r)$ and $b(r)$ for the geometry with a wormhole at $r=1$ and simple horizons at $r=r_0$.}\label{Fig-AbNeed}}\hfill
\parbox[t]{0.51\textwidth}{\caption{The global structure diagram of the model. The thick lines correspond to $r=\infty$ and the thin lines correspond to the horizons at $r=r_0$. }\label{Fig-CPdiag}}
\end{figure}

The horizon conditions correspond to simultaneous solutions of the
equations $b(r_0)=r_0$ and $A(r_0)=0$ as it is depicted in
Fig.~\ref{Fig-AbNeed}. We deal with only simple horizons. It means
that the metric component $g_{rr}$ has a simple pole at the point
$r_0$, so that
\beq \label{hrznCond1}
b(r)=r_0+b'(r_0)(r-r_0)+o(r-r_0)
\eeq
and the function $A(r)$ has a simple zero
\beq \label{simpleZero}
A(r)\propto (r-r_0)+o(r-r_0).
\eeq
These horizon conditions and  compatibility of the Einstein equations near the horizon lead to the following relation for the Chaplygin gas
\beq \label{hrznCond2}
b'(r_0)=8\pi r_0^2 \sqrt\alpha.
\eeq
In this case $r_0$ corresponds to the simple horizon separating a
static R~region at $r<r_0$ from a non-static  T~region at $r>r_0$. Indeed, for the metric (\ref{SSM}) the square
of the normal to a surface $r$=const is $\Delta =  b/r-1$
and $\Delta<0$ when $1<r<r_0$. In this region the spacetime is static (R~region) and $r$ is a spatial coordinate. At $r>r_0$ the normal is a spacelike vector and $\Delta>0$. This is a non-static T~region where $r$ becomes a temporal coordinate.

This class of metrics is regular in any finite region for $r\ge
1$. It is verified using the Kretschmann scalar
$K=R_{\alpha\beta\gamma\delta}R^{\alpha\beta\gamma\delta}$, where
$R_{\alpha\beta\gamma\delta}$ is the Riemann curvature tensor. To
avoid too cumbersome representation for the Kretschmann scalar we
write it in the form
\beq
K(r)=\frac{P(r,A,A',A'',b,b')}{A^4(r)r^6},
\eeq
where the numerator is a polynomial function depending on the
indicated arguments and $P\propto (r-r_0)^4 +
o\left([r-r_0]^4\right)$ near the horizon $r=r_0$. Taking into
account the expansion (\ref{simpleZero}) it means that the
Kretschmann scalar is finite on the horizon.

The metric (\ref{SSM}) in the T~region can be rewritten as
\beq \label{KSmetr}
\d s^2= \frac{\d\eta^2}{b(\eta)/\eta-1} - \tilde{A}(\eta)\d \rho^2 - \eta^2\d\Omega^2,
\eeq
where $\tilde{A}(r)=-A(r)$ is the positive function when $r>r_0$. We introduce the new coordinates, $\eta$ instead $r$ and $\rho$ instead $t$ to mark the interchange of roles between the spatial and temporal coordinates. The integral
\beq \label{PhysTime}
\tau = \pm\int \frac{\d\eta}{\sqrt{b(\eta)/\eta-1}}
\eeq
defines the physical time
$\tau$ for an observer in the T~region. The metric (\ref{KSmetr}) describes a Kantowski-Sachs (KS) anisotropic cosmological model \cite{KS} with two scale factors for the spherical and longitudinal directions which depend on the proper time $\tau$ according to $a_t(\tau)=\eta(\tau)$ and $a_r(\tau)=\tilde{A}^{1/2}\bigl(\eta(\tau)\bigr)$  correspondingly. The global structure of the spacetime is shown in Fig.~\ref{Fig-CPdiag}. The metric (\ref{KSmetr}) describes both the contracting region T$_-$ corresponding to $\tau<0$ and the expanding region T$_+$ corresponding to $\tau>0$.

At $\eta=r_0$ there is a coordinate singularity which the observer
interprets as the beginning of the universe. In \cite{Bronnikov2} it
is called  a \emph{null bang}  sinse a horizon is a null surface
where the spatial volume vanishes. It is easy to see from
Eq.~(\ref{PhysTime}) that if the function $b\to\infty$ faster than
$\eta^3$  for $\eta\to\infty$, then the universe evolves during a
finite cosmological time and otherwise it lives endlessly.

It is evident from the Einstein equations that the following conditions are fulfilled on the horizon
\beq \label{hrznCond3}
\rho(r_0)=-p_r(r_0)= \sqrt\alpha,
\eeq
that corresponds to the requirement $p_r+\rho=0$ on the  horizon
\cite{Bronnikov3} which is independent of the matter kind. The
condition (\ref{hrznCond3}) provides continuity of the energy
density and the radial pressure in going from R to T~region. In
fact, $\rho$ and $p_r$ interchange their positions in the
stress-energy tensor when crossing the horizon. The equation of
state (\ref{Chap}) is precisely the same in both regions. Under the
conditions (\ref{simpleZero}) and (\ref{hrznCond2}) the transversal
pressure is also a continuous function and takes the value
$p_t(r_0)=b''(r_0)/8\pi r_0-3\sqrt\alpha$ on the horizon.

Below some exact solutions of the equations
(\ref{Chap})--(\ref{Ein3}) with  the conditions
(\ref{throatCond1})--(\ref{hrznCond2})  corresponding to different
cosmological scenarios are studied.

\section{Different types of Kantowski-Sachs cosmologies}
\label{Sec-ExactSolutions}

The equation of state (\ref{Chap}) relates the energy density  with
the radial pressure only. It allows one to solve the problem
predetermining the metric function $b(r)$ satisfying the flare-out
and horizon conditions. The other functions are found from
Eqs.~(\ref{Ein1})--(\ref{Ein3}) and produce different cosmological
scenarios in the T~regions whereas the R~regions are similar in all
solutions.

\subsection{Eternally expanding universes}

One of the simplest functions satisfying the conditions (\ref{throatCond1}) and (\ref{hrznCond1}) is the parabola
\beq\label{b 2}
b(r)=d (r-1)(r-r_0)+r.
\eeq
The function $A(r)$ is expressed from Eq.~(\ref{Ein2}) with the conditions (\ref{throatCond2}) and (\ref{simpleZero}) as
\beq \label{A 2}
A(r)= A_0\frac{(r_0-r)(r+q_1)^{q_2}}{r}\,e^{q_3 r+q_4 r^2},
\eeq
where
\beq\label{params2}
\begin{array}{lcr}
&\displaystyle
q_1=\frac{r_0 d+d-1}{2d},\quad
q_2=\frac{2q_1^4}{r_0^4}\,\frac{1 - d + r_0 d }{ 1 + d - r_0 d},\quad
q_3=2q_4\left(3q_1+\frac{1}{d}\right),
&\\&\displaystyle
q_4=\frac{(r_0 d -d+1)^2}{4 r_0^4 d^2},\quad
d=\frac{r_0^2 \sqrt{r_0^4+8}- r_0^4-2 }{2 (r_0-1)}
&
\end{array}
\eeq
are positive constants and $r_0$ is connected with the parameter
$\alpha$ from the equation of state (\ref{Chap}) by the relation
\beq
\alpha=\frac{(1-d-r_0 d)^2}{64\pi^2 r_0^4}.
\eeq

\begin{figure}[t]
\includegraphics[width=.51\textwidth]{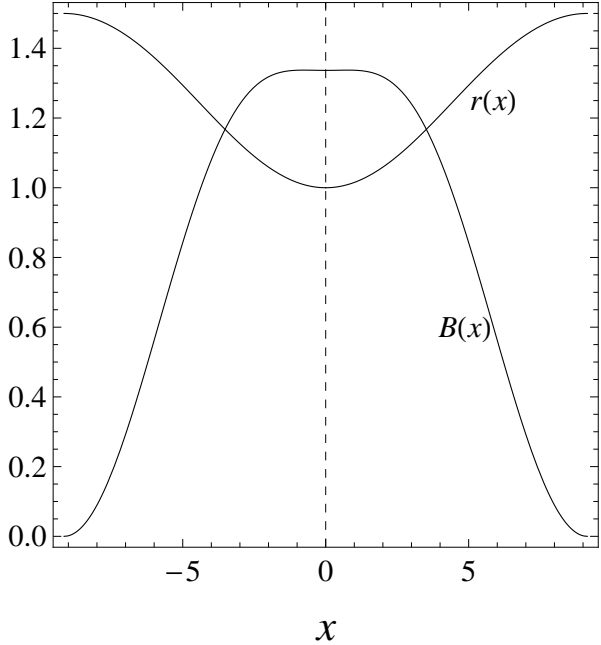}\\%
\parbox[t]{0.51\textwidth}{\caption{The metric functions depending on the Gaussian coordinates in the R~region. This figure is plotted for $b(r)$ given by Eq.~(\ref{b 2}) when $r_0=1.5$ and other parameters are calculated according to (\ref{params2}), however the functions $r(x)$ and $B(x)$ have a similar form for the other solutions. }\label{Fig-Rmetrics2}}
\end{figure}

In the R~region the metric (\ref{SSM}) can be transformed to the form
\beq\label{GuassMetr}
\d s^2 = B(x) \d t^2 - \d x^2 - r^2(x)(\d\theta^2+\sin^2\theta \d\varphi^2),
\eeq
 where
\beq\label{r in R 2}
r(x) = 1 + (r_0-1)\sin^2 \varepsilon\!\left(\sqrt d\, x/2\right),
\eeq
and $\varepsilon(z)$ is the inverse function to the elliptic integral of the second kind $E\bigl(\varepsilon \bigl|\bigr. 1-r_0\bigr)$. The metric function $B(x)=A(r(x))$, where $r(x)$ from Eq.~(\ref{r in R 2}) is substituted into the expression (\ref{A 2}). Behaviors of $r(x)$ and $B(x)$ are depicted in Fig.~\ref{Fig-Rmetrics2}.

\begin{figure}[t]
\includegraphics[width=0.42\textwidth]{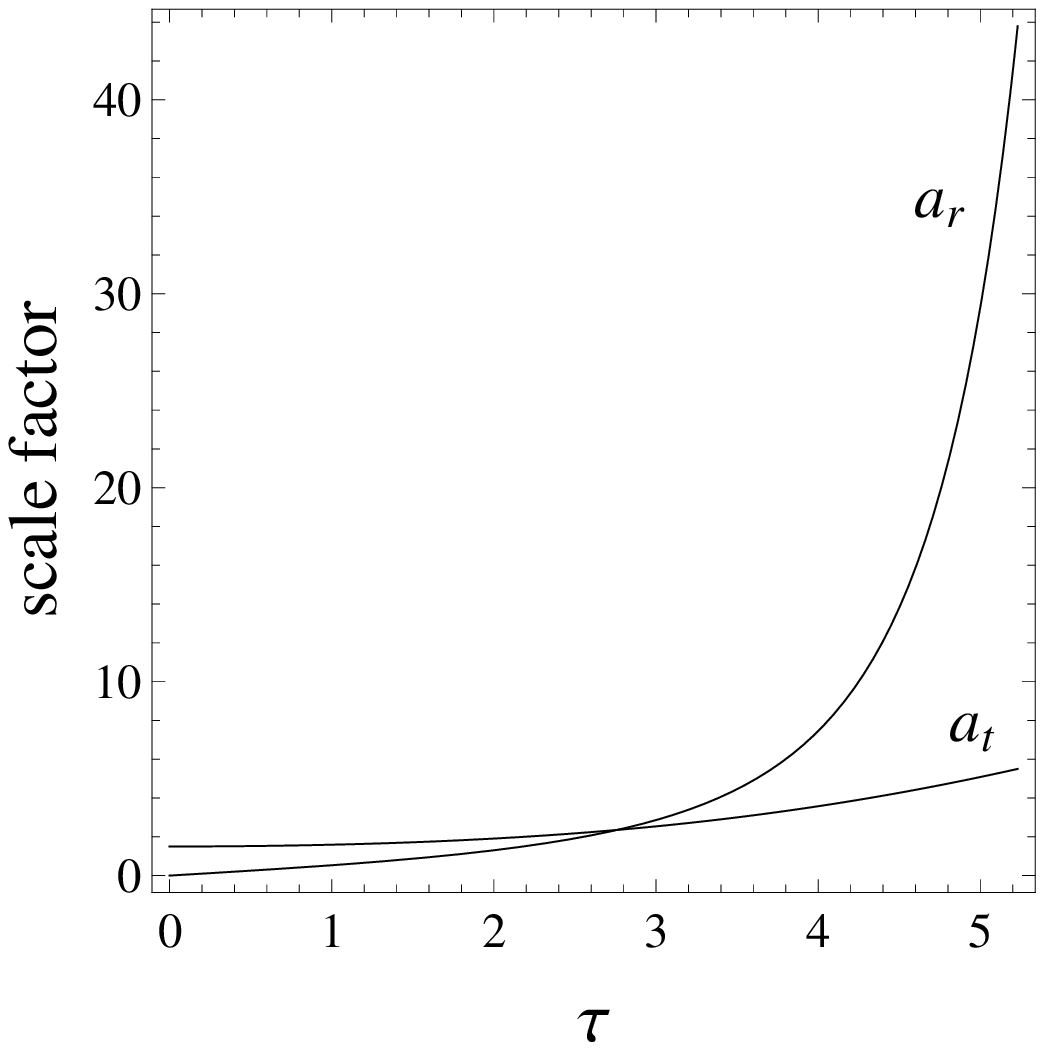}\hfill%
\includegraphics[width=0.52\textwidth]{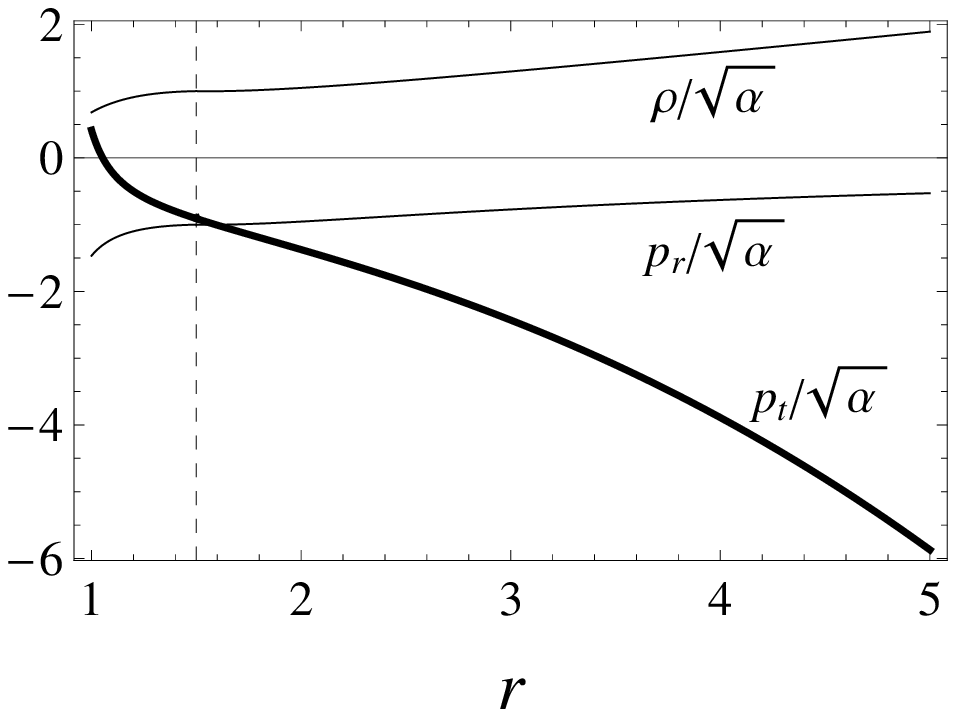}\\
\parbox[t]{0.42\textwidth}{\caption{The increase of the scale factors for the solution specified by Eqs.~(\ref{b 2})--(\ref{params2}) for  $r_0=1.5$. }\label{Fig-Tmetrics2}}\hfill
\parbox[t]{0.52\textwidth}{\caption{\label{Fig-prho2}
The dimensionless energy density, radial and transversal pressures as functions of the radial coordinate for the eternally expanding universe.}}
\end{figure}

Contrary to the representation (\ref{SSM}), the metric
(\ref{GuassMetr}) describes the whole R~region with the Gaussian
radial coordinate $x$ lying in the range $|x|\le\pi E\bigl(\pi/2 \bigl|\bigr. 1-r_0\bigr)/\sqrt d$.
When $x\to\pm\pi E\bigl(\pi/2 \bigl|\bigr. 1-r_0\bigr)/\sqrt d$ the function $B(x)$ becomes zero
providing the evidence for horizons which are spaced at a finite
distance apart for a static observer. The function $r(x)$ has a
minimum at $x=0$ implying a wormhole throat, so the
R~region is a spherically symmetric one with the wormhole
surrounded by two horizons.

For observers in the T~region the universe evolution starts from the horizon at some instant in the past and lasts for  infinite time. All this time the universe expands in all directions as it demonstrates in Fig.~\ref{Fig-Tmetrics2}.
In the far future the scale factor along the coordinate spheres is quadratic in the proper time $\tau$  while the scale factor in the radial direction grows exponentially
\beq
a_r\propto \tau^{q_2}\exp q_4\tau^4/2,\quad a_t\propto \tau^2.
\eeq

The behaviors of the energy density and the pressures are shown in
Fig.~\ref{Fig-prho2}. The energy density is positive and the radial
pressure is negative in the whole spacetime whereas the transversal
pressure reverses its sign. In the T~region all these quantities are
comparable  close to the null bang. At the late stage the radial
pressure decreases in its absolute  value as $p_r\propto\tau^{-2}$
and the main contribution is given by the energy density growing
quadratically in the proper time, $\rho\propto\tau^2$,  and the
negative transversal pressure which increases in its absolute value
as $p_t\propto\tau^6$.

\subsection{Universes with finite lifetime}

The alternative scenario is arrived from the equations (\ref{Chap})--(\ref{Ein3}) if the function  \beq\label{b 4}
b(r)=dr^4+1-d.
\eeq
In this case the function
\beq\label{A 4}
A(r)= A_0\frac{(r_0-r)}{r\bigl[ r^2+(r_0+1)r+q_1^2q_2+r_0
\bigr]^{q_1}}\exp\left\{-\frac{4r_0q_1}{q_2} \,
\text{arctan}\left(\frac{2r+r_0+1}{q_2}\right)\right\},
\eeq
where
\beq
q_1=\frac{r_0^2+1}{q^2},
\qquad
q_2=\sqrt{3r_0^2+2r_0+3},
\qquad
r_0=1+\sqrt 2.
\eeq
This solution implies $\alpha=(3 - 2\sqrt 2)/64 \pi^2$ in the equation of state (\ref{Chap}).

In the R~region the spacetime has the same structure as before. The
metric can be transformed to the form (\ref{GuassMetr}) with the
only difference that the metric functions $B(x)$ and $r(x)$ cannot
be obtained analytically and all calculations are numerical. Their
qualitative behaviors are the same as in Fig.~\ref{Fig-Rmetrics2}

\begin{figure}[t]
\includegraphics[width=0.42\textwidth]{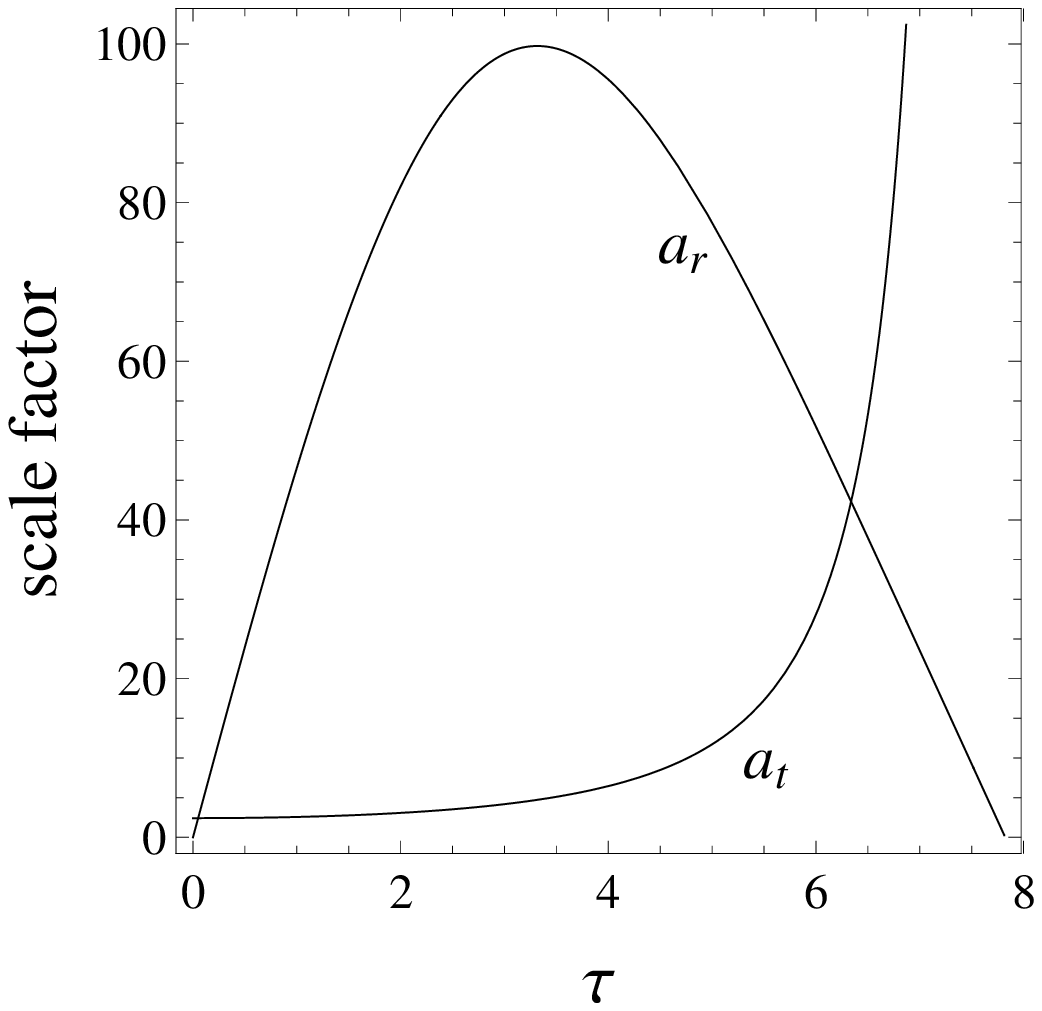}\hfill%
\includegraphics[width=0.52\textwidth]{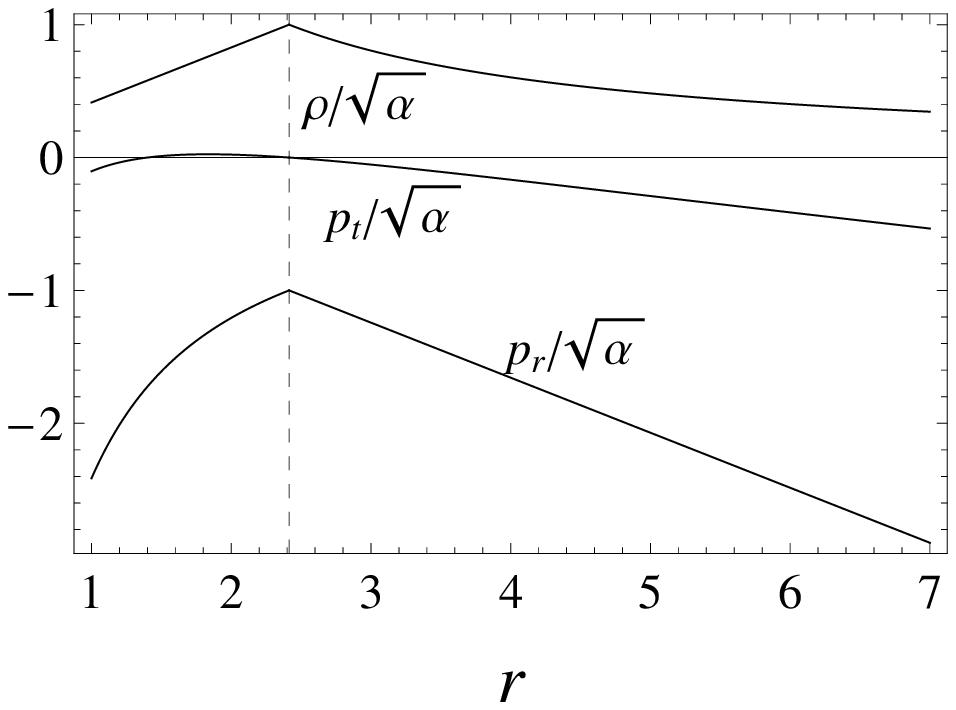}\\
\parbox[t]{0.42\textwidth}{\caption{The behaviors of the scale factors for the solution  when the function $b(r)$ and $A(r)$ are given by Eqs.~(\ref{b 4}) and (\ref{A 4}). The proper lifetime of the universe is finite.}\label{Fig-Tmetrics4}}\hfill
\parbox[t]{0.52\textwidth}{\caption{The energy density, radial and transversal pressures as a functions of the radial coordinate in the universe with the finite proper lifetime.}\label{Fig-prho4}}
\end{figure}

The expansion in the T~region starts with a null bang and the
ensuing cosmological evolution develops during finite physical
lifetime from the point of view of the KS observer.
Fig~\ref{Fig-Tmetrics4} demonstrates that throughout this period the
spherical scale factor infinitely increases and the longitudinal
scale factor peaks at some instant and falls to zero, although both
radial and transversal pressures are negative and the energy density
disappears  towards the end of the evolution as it is shown in
Fig.~\ref{Fig-prho4}. This situation arises  in the KS model from
the fact that a transversal acceleration is determined by the radial
pressure and the transversal Habble parameter $H_t=\dot a_t/a_t$,
where the dot denotes the derivative with respect to the proper time
$\tau$, while a radial acceleration also depends on the transversal
pressure and the energy density according to
\beqar
&\displaystyle
\frac{1}{a_t^2}+H_t^2+2\frac{\ddot a_t}{a_t}=-8\pi p_r,
&\\&\displaystyle\label{radialAcceleration}
\frac{1}{a_t^2}+H_t^2-\frac{\ddot a_r}{a_r}=4\pi \left( \rho -p_r +2p_t\right).
&
\eeqar
If the energy density  is neglected and the rate of growth of $H_t$
is small so that $|p_r|>2|p_t|+H_t^2/4\pi$, then one observes a
decelerated expansion or  contraction  in the longitudinal
direction.

\subsection{Eternally contracting universes}

An intermediate scenario is obtained for the metric functions
\beqar
&\displaystyle\label{b 3}
b(r) = dr(r-1)(r-r_0)+r,
&\\ &\displaystyle\label{A 3}
A(r) = A_0\frac{(r_0-r) }{r^{q_3}\bigl[3 r^2 \!\!-\! 2q_1r \!\!+\! r_0q_3 \bigr]^{q_2q_5}}\exp\left\{-2\frac{q_1q_4q_5}{q_6} \text{arctan}\left(\frac{3r -q_1}{q_6} \right)\right\},
&
\eeqar
where
\beq\label{params3}
\begin{array}{lcr}
&\displaystyle
q_1=r_0+1 ,
\quad
q_2=d(2+5r_0+2r_0^2)-2d^2r_0(1+r_0+r_0^2)-1,
&\\&\displaystyle
q_3=1+\frac{1}{dr_0}  ,
\quad
q_4=d(2+5r_0+2r_0^2)-d^2r_0(2-r_0+2r_0^2)-4,
&\\&\displaystyle
q_5=\frac{1-dr_0+dr_0^2}{6d^2r_0^4(1+d-dr_0)},
\quad
q_6=\sqrt{\frac{3}{d}-1+r_0-r_0^2},
&\\&\displaystyle
\alpha=\frac{(1-dr_0+dr_0^2)^2}{64\pi^2 r_0^2},
\quad
d=\frac{r_0\sqrt{r_0^4+4r_0^2+4r_0}-r_0^3-2}{2(r_0^2-1)}
&
\end{array}
\eeq
are positive constants except $q_2$, which can be both positive and
negative\footnote{Nevertheless the value $q_3+2q_2q_5-1>0$ for all
values of $r$ lying in the range $1<r_0<r_\text{max}$. It means that
$A(r)$ vanishes when $r\to\infty$.}, and the horizon radius lies in
the range $1<r_0<r_\text{max}\approx3.04$. If $r_0\ge r_\text{max}$
the function $A(r)$ resulting from Eq.~(\ref{Ein2}) gets
singularities in the static region.

Note that the transformation to the metric  (\ref{GuassMetr})  in
the R~region gives the quite simple form for spherical radii
\beq
r(x) = \frac{1}{2}\left((r_0+1) -(r_0-1)\cos\sqrt d x\right)
\eeq
with  the Gaussian coordinate  $x$ lying in the range $[-\pi/\sqrt d,\pi/\sqrt d]$.

\begin{figure}[t]
\includegraphics[width=0.42\textwidth]{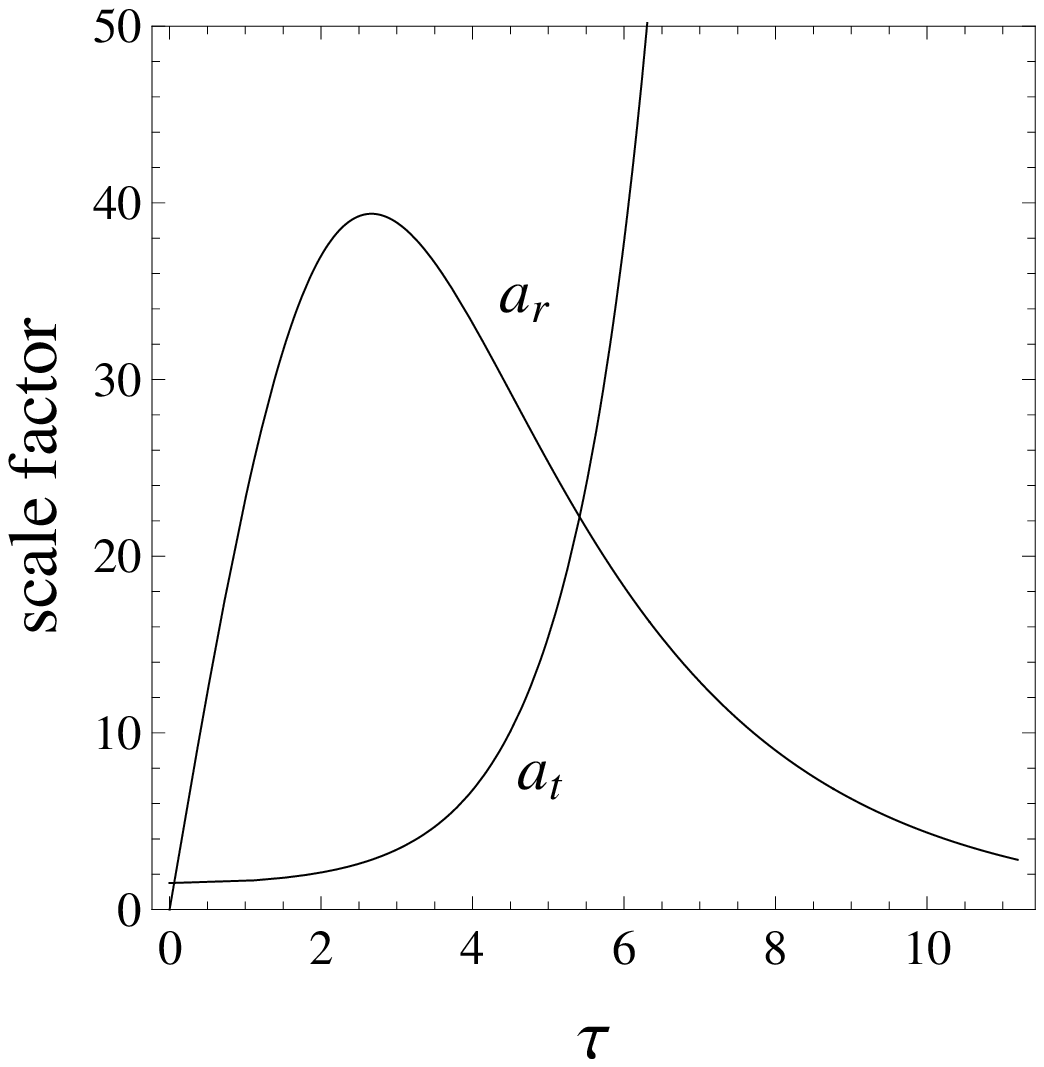}\hfill%
\includegraphics[width=0.52\textwidth]{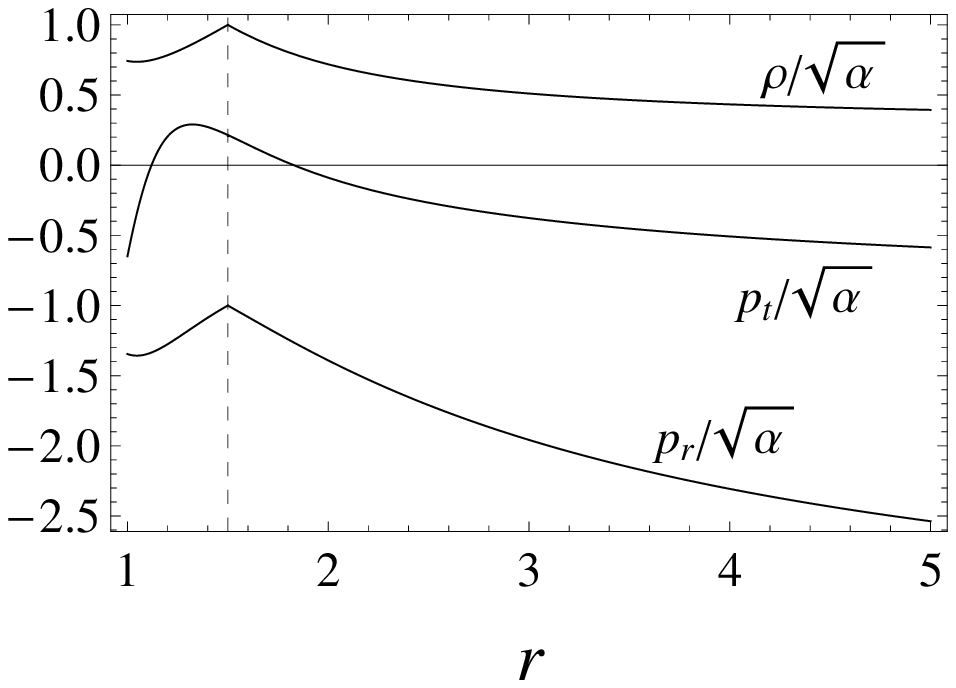}\\
\parbox[t]{0.42\textwidth}{\caption{The behaviors of the scale factors in the solution with the metric function (\ref{b 3}) and (\ref{A 3}) where the quantities (\ref{params3}) are calculated for $r_0=1.5$. The universe expands in the spherical direction and contracts in the radial direction during an infinite time.}\label{Fig-Tmetrics3}}\hfill
\parbox[t]{0.52\textwidth}{\caption{The energy density, radial and transversal pressures as functions of the radial coordinate in the gradually dying universe.}\label{Fig-prho3}}
\end{figure}

For the KS observer the universe starts to evolve  from the horizon
and  gradually dies into infinite future because  the expansion in
the radial direction gives place to the contraction. The scale
factors for the longitudinal and lateral directions as functions of
the proper time are shown in Fig.~\ref{Fig-Tmetrics3}. The radial
pressure is everywhere negative (see Fig.~\ref{Fig-prho3}), so that
it guarantees the accelerated expansion along the coordinate
spheres, and the inequality $\rho-p_r+2p_t-H_t^2/4\pi>0$ provides
the decelerated growth and the subsequent decay of the radial scale
factor according to Eq.~(\ref{radialAcceleration}).  At the late
stage the scale factors vary in the proper time $\tau$ by the
exponential laws
\beq \label{almostdS}
a_r\propto\exp\left\{-\sqrt d \left(q_2q_5-\frac{1}{2dr_0}\right)\tau\right\},
\quad
a_t\propto\exp\sqrt d \,\tau.
\eeq
The asymptotic expression (\ref{almostdS}) resembles in some way the de Sitter behavior bringing the pressures and the energy density to the constant values
\beq
\rho\to\frac{8\pi\alpha}{3d},\quad
p_r\to -\frac{3d}{8\pi},\quad
p_t\to -\frac{d}{32\pi}\left[ 4+(3-q_1-2q_2q_3)(1-q_1-2q_2q_3)\right].
\eeq
Such a tendency takes place for any function $b(\eta)$  increasing
as $\eta^3$ when $\eta\to\infty$. It can provide the basis for a
cosmological scenario with an asymptotic de Sitter  phase at late
times.

\section{Conclusion}
\label{Sec-Conclusion}

We have considered some exact solutions of Einstein's equations with
anisotropic fluid describing a spacetime with two identical
cosmological T~regions and an intermediate static spherically
symmetric R~region. The latter contains a wormhole allowing one to
pass between two horizons. The T~regions are described by
anisotropic KS models with different kinds of cosmological
evolution. We have presented the solutions describing eternally
expanding universes, disappearing ones and universes with a limited
lifetime.

These solutions rest upon the flare-out  conditions
(\ref{throatCond1}), (\ref{throatCond2}) and the horizon regularity
conditions for matter with the Chaplygin equation of state
(\ref{hrznCond1})--(\ref{hrznCond2}), (\ref{hrznCond3}). Similar
regularity conditions were derived before for the null bang
cosmologies with different types of matters such as the neutral and
charged matter with a barotropic equation of state $p_r\sim \rho^m$,
$m\ge 1$ \cite{Bronnikov4} and matter with a non-vacuum behavior
(i.e. the ratio of the radial pressure to the energy density
$w=p_r/\rho\ne -1$ everywhere) \cite{Bronnikov3}. However they are
incompatible with the model under consideration since the regularity
for the Chaplygin gas is realized only through the ``vacuum'' value
$w=-1$ on the horizon.

The presented solutions violate everywhere  the dominated energy
condition, $\rho>0,\ \rho>|p_r|,\ \rho>|p_t|$ while the strong
energy condition proves to be realized in some parts of the
spacetime as it takes place in the last scenario. One sees from
Fig.~\ref{Fig-prho3} that the relation $\rho+p_r+2p_t>0$ holds  in
the layer near the horizon.

Note that the suggested spacetime  conformation does not support an
isotropic solutions in both R~and T~regions. Indeed, the isotropy of
a spacetime implies $p_r= p_t$ and Eq.~(\ref{Ein3}) is readily
integrated. The resulting function $A(r)$ vanishes concurrently with
the energy density and hence with the derivative $b'(r)$.
Fig.~\ref{Fig-AbNeed} pictorially shows that such a solution is
incompatible with the considered structure.

Besides, there  is no an isotropization in the solutions obtained,
although it has considerable utility for the early Universe model.
Nevertheless, isotropization can be most likely realized in the
studied model. The last scenario has the asymptotically constant
pressures and energy density, and solutions of this kind  may quite
probably  demonstrate the de Sitter behavior at late times for the
special selection of the metric functions. This problem is the
subject of our current studies and the  results will be presented in
our future papers.

\end{document}